# Unconventional Geometric Phase in Twisted Junction of Cuprates

Zhigang Song[1], Kai Chang[1,2]

1Center for Quantum Matter, School of Physics, Zhejiang University, Hangzhou 310058, China

2 Institute for Advanced Study in Physics, Zhejiang University, Hangzhou 310058, China

Originally introduced in optics, the Pancharatnam-Berry phase is a general concept of geometric phase defined for any two interfering polarization states. In electronic systems, however, its counterpart has long been overlooked due to the absence of electron polarization. Here, using large-scale first-principles calculations, we investigate the electronic structure of twisted bilayer $Bi_2Sr_2CaCu_2O_8$. We find spontaneous spin polarization and the emergence of hidden flat bands at the interface between atomic layers. Most notably, we discover an unconventional geometric phase analogous to the Pancharatnam-Berry phase in optics. This electronic geometric phase exerts opposite effects on superconducting currents of opposite chirality, enabling twisted cuprates to act as a filter for chiral superconducting current—even if the ground state itself is non-chiral.

Correspondence: Kai Chang (kchang@zju.edu.cn) or Zhigang Song (szg@pku.edu.cn)

The Berry phase is a specific type of geometric phase that arises from adiabatic and cyclic evolution in parameter space. In electronic systems, Berry-phase effects underlie several important phenomena, including the anomalous Hall effect, topological insulators, valley Hall effects in two-dimensional materials, orbital magnetization, and electric polarization.[1] Originating from optics, the Pancharatnam-Berry phase represents a more general form of geometric phase. It is defined for any two interfering polarization states and does not strictly require a closed loop in parameter space.[2] Unlike photons, electrons in solids generally lack a well-defined spatial polarization analogous to photon polarization. Moreover, due to the $U(1)$ gauge symmetry of electron wavefunctions, any phase accumulated along an open path can be removed via gauge transformation. As a result, Pancharatnam-type geometric phases have long been overlooked in electronic systems. However, in superconductors with unconventional pairing symmetry, such as $d$-wave or $p$-wave, the Cooper pairs can exhibit a spatial polarization akin to that of photons. To date, the presence of a Pancharatnam-type geometric phase and its effects on electronic states has rarely been reported.

In moiré superlattices formed by twisted bilayer materials, the interplay between geometric phases and strong electronic correlations has led to a variety of emergent collective phases,[3-5] such as fractional Chern insulators, orbital ferromagnetism,[6] and the possible topological superconductivity.[7] The polarization associated with unconventional superconductivity and the twisted geometry make moiré cuprate systems a promising platform to explore Pancharatnam-type geometric phase. In this work, we carried out large-scale charge density functional theory (DFT) calculations[8-10] to investigate the electronic structure of twisted bilayer $Bi_2Sr_2CaCu_2O_8$, serving as a model system for twisted cuprate junctions. Our calculations revealed a Pancharatnam-type geometric phase in the electronic states. This unconventional phase exerts opposite effects on superconducting current of opposite chirality, allowing twisted cuprates to act as filters for chiral superconducting states—even if the ground state itself is not chiral.

**Numerical calculations**

When two $Bi_2Sr_2CaCu_2O_{8+x}$ layers are twisted by a mutual angle close to $45°$, theoretical studies have predicted a fully gapped chiral superconducting phase with broken time-reversal symmetry.[11-16] Experimentally, moiré superlattices of $Bi_2Sr_2CaCu_2O_{8+x}$ with various twist angles have recently been fabricated. Although some controversy remains due to differences in sample fabrication conditions, angular dependencies consistent with a dominant $d$-wave order have been observed, particularly in $45°$-twisted $Bi_2Sr_2CaCu_2O_{8+x}$ layers in experiments.[17-19] As the twist angle approaches $45°$, the Josephson critical current decreases by up to two orders of magnitude compared with untwisted Josephson junctions [20, 21]. Experimental signatures of time-reversal symmetry breaking, including reversible Josephson diode behavior, have also been reported.[19] In theory, prior work has emphasized the importance of interlayer coupling in modeling these systems, yet most analyses rely on assumed interlayer parameters.[7] To date, the fundamental electronic structure of twisted bilayer cuprates has not been reported via DFT calculation, largely due to the computational challenges posed by the large number of atoms in the moiré unit cell.

We started our work by performing large-scale DFT calculations to determine the electronic structure of twisted bilayer $Bi_2Sr_2CaCu_2O_8$. We employed a single-zeta atomic basis set and the PBE exchange-correlation functional, using FHI pseudopotentials. To incorporate interlayer van der Waals interactions, we applied the DFT-D2 method,[22] and relaxed the atomic positions until the residual force on each atom was below 0.03 eV/Å. A $3\times3\times1$ $k$-point mesh was used for Brillouin zone sampling. To account for possible strong electron correlations, we performed GGA+U calculations with varying Hubbard U values applied to the Cu $d$ orbitals. Test results (Fig. S1) indicate that the choice of $U$ does not significantly affect the key conclusions. Therefore, we set $U = 0$ in subsequent calculations.

The relaxed structures at different twist angles exhibit a consistent physical behavior. An example at a twist angle of 43.6° is shown in Fig. 1(a). The two Cu–O planes are well separated—by more than 1 nm—and do not come into direct contact. The relaxed structure is in good agreement with previous experimental results, as illustrated in Fig. 1(a, b). Although the Cu–O planes experience slight distortions, they remain largely intact. In contrast, significant structural deformation occurs at the interface layer, primarily involving the Bi–O planes, as shown in Fig. 1(c). The oxygen atoms, in particular, undergo substantial displacement, leading to the formation of a disordered array of Bi–O dimers. Due to their small atomic radius, oxygen atoms are highly sensitive to the local environment and do not occupy fixed positions, resulting in random, localized distortions. Consequently, electrons in the Bi–O layers experience a strongly modulated moiré potential after twisting. Notably, similar large distortions of the Bi–O layers have been discussed in untwisted $Bi_2Sr_2CaCu_2O_{8+x}$ in earlier theoretical studies.[23]

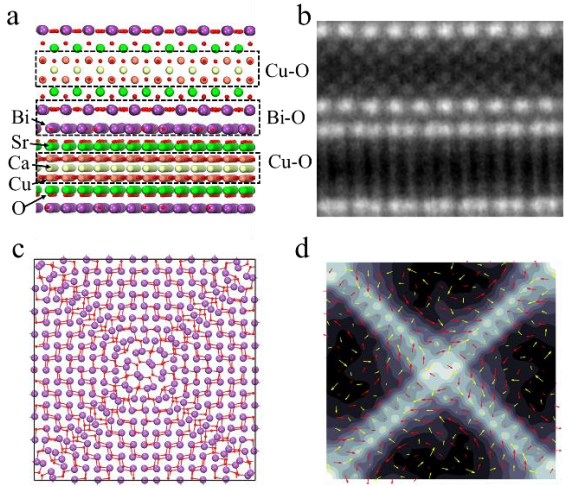

**Figure 1** (**a**) Relaxed atomic structure of twisted bilayer $Bi_2Sr_2CaCu_2O_8$ at a mutual twist angle of 43.6°. (**b**) High-resolution TEM image reproduced from Ref [21]. (**c**) Top view of the Bi–O interface layers in twisted bilayer $Bi_2Sr_2CaCu_2O_8$ at a twist angle of 6.03°. (**d**) Atomic displacements of Bi (red) and O (yellow) atoms after relaxation in one Bi–O plane. The color scale indicates the displacement along the $z$-axis.

The band structures of twisted bilayer $Bi_2Sr_2CaCu_2O_8$ at various twist angles are shown in Fig. 2(f), and for more structures with different twisted angles are shown in Fig. S2. Near the Fermi level, the bands are highly entangled. At small twist angles, the band structure becomes increasingly incoherent, as shown in Fig. S3. This is a behavior distinct from that observed in twisted bilayer graphene or other twisted semiconductors.[24-28] The Cu-O planes undergo only slight distortion, and the bands projected onto the Cu-$d_{x2-y2}$ orbitals remain largely unaffected. These bands are approximately folded versions of those in the untwisted structure and retain significant dispersion. Across different twist angles, the energy perturbation at high-symmetry points ranges from 1 to 20 meV. This implies that the supermodulation arising from the direct interlayer coupling of Cu-O layers and the indirect interaction due to the distortion of Bi-O layers is not large.[29]

If we isolate the intralayer distortions, the Cu-$d_{x^2-y^2}$ projected bands in separated layers are nearly identical to those in the twisted bilayer, as shown in Fig. S2. This indicates that direct coupling between the Cu–O planes is negligible. In contrast, the bands originating from the Bi–O planes are strongly distorted near the Fermi level. These bands are nearly flat, with bandwidths below 50 meV. As the twist angle decreases and the moiré lattice constant increases, both the bandwidth and the energy separation between bands further decrease. Although much attention has been focused on the Cu-$d_{x^2-y^2}$ bands due to their relevance in high-temperature superconductivity, the flat bands associated with the Bi–O layers are often overlooked—particularly in simplified theoretical models of twisted cuprates.

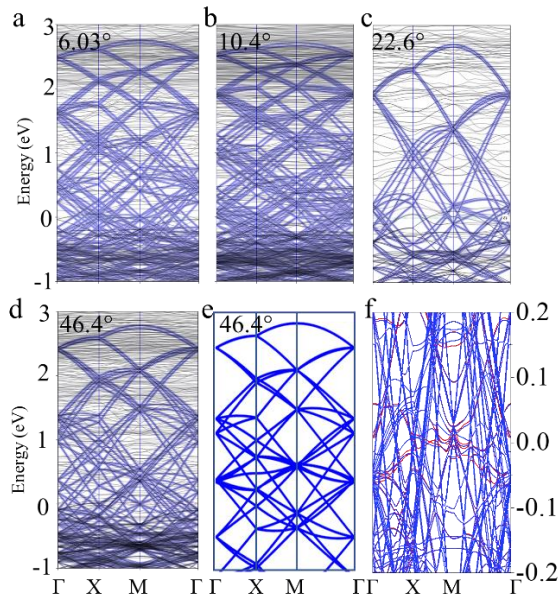

**Figure 2** (**a-e**) Orbital-projected band structures of twisted bilayer Bi$_2$Sr$_2$CaCu$_2$O$_8$ at various mutual twist angles. The weight represents the projection onto Cu-$d_{xy}$ and Cu-$d_{x^2-y^2}$ orbitals. (**e**) Band structure calculated using a tight-binding model with a nearest-neighbor hopping integral set to –0.6 eV. (**f**) Spin-polarized band structure of the 46.4° twisted bilayer, showing spontaneous spin splitting.

In our calculations, we included temperature effects via Fermi-Dirac smearing. Interestingly, we observed that the spin degeneracy of the flat bands is lifted as the temperature decreases, indicating spontaneous breaking of time-reversal symmetry as shown in Fig. S4. The spin splitting reaches approximately 10 meV at various twist angles when the temperature is lowered to 100 K. However, this spin splitting and the magnetic moments—likely induced by doping—disappears at temperatures above 250 K. As the twist angle decreases, the bands near the Fermi level become increasingly flat, which enhances the spin splitting and raises the critical temperature for magnetic ordering. This kind of spontaneous ferromagnetism in partially occupied flat bands has been predicted in theory[6] and observed in recent experiments.[30] Notably, the experimentally observed superconducting transition temperature (~80 K) is lower than the critical temperature for the onset of magnetism predicted in our calculations. The spin polarized is mainly projected on the Bi–O layers. In previous studies, the broken time-reversal symmetry was usually attributed to $d+id$ superconducting order[16, 31] or loop currents[32-35].

**Model Analysist**

To avoid the limitations of DFT in treating strong electron correlations and pairing interactions, we study the properties of twisted bilayer cuprates within a mean-field framework. Based on our numerical results, the Hamilton of twisted bilayer $Bi_2Sr_2CaCu_2O_8$ can be written as

$$H = \begin{pmatrix} H_S(\frac{\theta}{2}) & H_{TS}(\mathbf{q}) & 0 \\ H_{TS}^{\dagger}(\mathbf{q}) & H_T & H_{TS}(-\mathbf{q}) \\ 0 & H_{TS}^{\dagger}(-\mathbf{q}) & H_S(-\frac{\theta}{2}) \end{pmatrix} \tag{1}$$

where $H_S(\frac{\theta}{2}) = \begin{pmatrix} -\xi\ (R_{\theta/2}\mathbf{k}) & \Delta(R_{\theta/2}\mathbf{k}) \\ \Delta^{\dagger}(R_{\theta/2}\mathbf{k}) & \xi\ (R_{\theta/2}\mathbf{k}) \end{pmatrix}$ is the Hamiltonian for a superconducting Cu-O layer. $R_{\theta/2}$ represents the twisting operation by an angle of $\theta\,/\,2$. $\Delta(\mathbf{k}) = \Delta_0(\cos(k_x) - \cos(k_y))$ is the $d$-wave superconducting order. The amplitude of quasiparticle gap $\Delta_0$ is ~40 meV.[36]

$\xi\ (R_{\theta/2}\mathbf{k}) = \sum_{\mathbf{k}} 2t(\cos(R_{\theta/2}k_x) + \cos(R_{\theta/2}k_y)) - \mu$, where $\mu$ is chemical potential. By fitting DFT band structures, $t$ is determined to be -0.6 eV. $H_T$ represents the sub-Hamiltonian corresponding to the tunneling layer composed of Bi and O atoms. The electronic structure of this tunneling layer is highly complex, and its detailed band features are difficult to extract directly. However, even without explicitly modeling the full tunneling layer, essential physical characteristics can still be captured, as indicated by our DFT results. Due to the strong structural distortion and complexity of the Bi–O interface, it is challenging to precisely determine the interlayer tunneling terms. $H_{TS}$ or $H_T$ and can be neglected, the resulting band structure derived from Eq. (1) approximates the DFT orbital-projected ("fat") bands shown in Fig. 2(e) and Fig. S2. Importantly, despite uncertainties in interlayer coupling, the key physics remains evident. The large moiré potential resulting from distortion in the Bi–O planes leads to flat bands in the Bi-O layers. These flat bands are susceptible to spontaneous spin splitting via Stoner-type ferromagnetic instability [6], which is consistent with our DFT findings.

The coupling between the superconducting Cu–O layers is indirect in nature, and it can be regarded as a type of superexchange interaction. In this case, model is reduced to a four-band model used in previous work.[7, 11] To estimate the average magnitude of the interlayer coupling constant $g$, we analyzed the energy splittings at high-symmetry points in the Brillouin zone, using them to approximate the effective coupling felt by Cu-$d_{xy}$ and $d_{x2-y2}$ electrons. The results are shown in Fig. 3(a). The estimated interlayer coupling $g$ ranges from approximately 1 meV to 20 meV as the twist angle increases from 4.58° to 46.4°. In general, $g$ decreases with decreasing twist angle for angles below ~36°. The calculated single-particle band structures are shown in Fig. S5.

The spin polarization is mainly located in the Bi-O layer, and thus it does not induce the electron-electron pairing. The spin polarization only induces an external phase difference between the upper and lower superconducting layer. In our tight-binding model, the interlayer superexchange coupling is approximated using a constant coupling $g$. For a twist angle of $46.4°$, our numerical calculations show that the total free energy is approximately $0.62$ meV lower in a chiral ground state with $d_{x2-y2} \pm i2d_{xy}$ symmetry, compared to a non-chiral superconducting state of the $d_{x2-y2}+2d_{xy}$ form. A comparison of the quasiparticle spectra for these two pairing states is presented in Fig. 3(b). Although the chiral superconducting state is the ground state, the barrier between non-chiral and chiral states is very low, and the coupling between $d_{x2-y2}$ and $d_{xy}$ states in the upper and lower layers is very weak. Given their potential applications,[37, 38] we focus on the interlayer phase difference and the interlayer tunneling current of a chiral superconducting state.

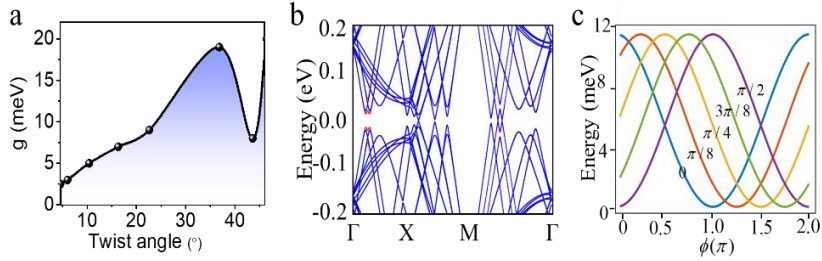

**Figure 3** (**a**) Estimated interlayer coupling strength $g$ as a function of twist angle. (**b**) Comparison of quasiparticle spectra for the decoupled non-chiral superconducting state ($d_{x2-y2} +2d_{xy}$ shown in blue) and the chiral superconducting state ($d_{x2-y2}+2id_{xy}$ shown in red). (**c**) Free energy as functions of applied phase. Different color represents different twist angles.

The effective Hamiltonian for each superconducting layer includes a self-energy term $H_{eff} = H_L(\frac{\theta}{2}) - \sum$ with self-energy term of $\sum = -H_{RT}[EI - H_T - H_{LT}^{\dagger}(EI - H_L(-\frac{\theta}{2}))^{-1}H_{LT}]^{-1}H_{TR}^{\dagger}$. $I$ is the identical matrix, and $E$ denotes the quasiparticle energy. Compared to the broad bandwidth of the Cu-$d_{xy}$ bands (~4000 meV), and the self-energy $\Sigma$ is relatively small. This correction originates from both intra-layer relaxation and indirect interlayer coupling, and is of the same order of magnitude as the estimated interlayer coupling strength $g$. This suggests that the two superconducting Cu–O layers can be treated as approximately layer independent degrees of freedom.

In each layer, the corresponding wavefunction can be mathematically decomposed into two chiral components. The upper and lower layers can therefore host wavefunctions of the form $\left| R_{\pm\theta/2}d_{xy} \right\rangle = \frac{1}{2i}(e^{-i\frac{\theta}{2}}\left| l_z = 2 \right\rangle - e^{i\frac{\theta}{2}}\left| l_z = -2 \right\rangle)$ for the upper layer and the lower layer, respectively. In the following part, we focus on the chiral degree of freedom. We first analyze each chiral component independently. For a given chirality, the low-energy effective Hamiltonian in a single layer can be expressed as:

$$H_S = \begin{pmatrix} -\xi\left(\mathbf{k}\right) & e^{-i\gamma}\Delta_0 e^{il_z(\varphi)} \\ e^{i\gamma}\Delta_0^{\dagger} e^{-il_z(\varphi)} & \xi\left(\mathbf{k}\right) \end{pmatrix} \quad (2)$$

where $l_z$ is the out-of-plane angular momentum quantum number associated with the pairing symmetry. In the case of cuprates, $l_z$ is $\pm 2$. The corresponding BCS ground-state wavefunction for the $d$-wave superconductor is:

$$\left|\psi_L\right\rangle = \prod_{\mathbf{k}}(u_{\mathbf{k}} + e^{-i\gamma}v_{\mathbf{k}}c_{\mathbf{k}\uparrow}^{\dagger}c_{-\mathbf{k}\downarrow}^{\dagger})\left|0\right\rangle \quad (3)$$

Here, the phase factor $\gamma$ is usually considered negligible under typical conditions. However, a key feature of expression (3) is that it does not conserve particle number, which makes the phase factor $e^{i\gamma}$ physically meaningful, even if $\gamma$ originates from a geometric phase such as the Pancharatnam-Berry phase.[39] When one layer undergoes an adiabatic twist by an angle $\theta$, the superconducting wavefunction acquires a global geometric phase as a result of this chiral structure.

$$\gamma = -i\int_0^{\theta/2}\left\langle\Psi\left|\frac{\partial}{\partial\varphi}\right|\Psi\right\rangle d\varphi = l_z\theta/2 \quad (4)$$

Such a phase is geometric in nature and represents an intrinsic property of the material system. The opposite superconducting layer acquires a corresponding phase of $-l_z\theta/2$. This phase is analogous to the Pancharatnam-Berry phase known from optics. In the supporting information, we present an independent method that arrives at the same conclusion, reinforcing the physical validity of this geometric phase (as shown in the supporting materials).

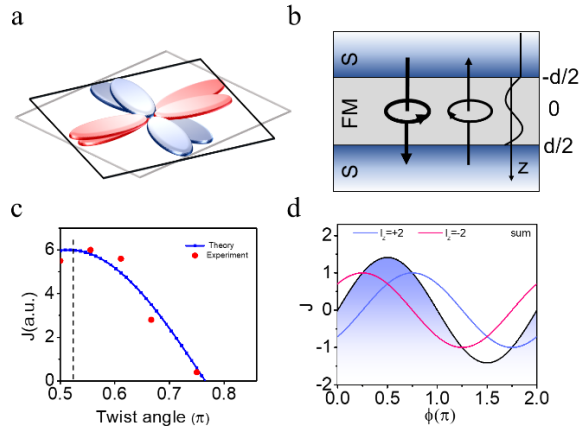

**Figure 4** (**a**) Illustration of twisted $d$-wave superconductivity. (**b**) Schematic illustration of chiral tunneling current between two superconducting Cu–O layers mediated by the Bi–O interface. (**c**) Comparison between theoretical and experimental angular dependence of Josephson current. The experimental data is measured at 5 K and obtained from ref.[21]. The current peak is located at $9\pi/20$ by fitting experiment. (**d**) Josephson current $J$ as a function of the superconducting phase difference $\phi$.

In single-particle physics, global phases are typically considered unobservable due to the underlying $U(1)$ gauge symmetry. However, in superconductors, such global phases can manifest in measurable quantities in Josephson junctions. The geometric phase modifies the Josephson current, which is generally expressed as $J \propto J_c \sin(\phi)$, where $\phi$ is the phase difference between the superconducting condensates. The presence of a geometric phase effectively shifts this relation to $J \propto J_c \sin(\phi + 2\gamma)$, demonstrating its direct impact on macroscopic observables. On another side, the numerical calculations based on the model show the free energy ($F$) is in the format of $F = F_0 + F_1 \cos(\phi + lz\theta)$ (as shown in Fig. 3(b)), $F_0$ is a constant and $F_0 = 0.62$ meV. The phase-current relation can be obtained by $J = \frac{2e}{\hbar} \frac{\partial F}{\partial \phi}$. The resulting current is $J = J_0 \sin(\phi + lz\theta)$. This is in accordance with our analysis.

To provide a more rigorous confirmation, we arrive at the same conclusion using a semi-empirical scattering model, informed by our DFT calculations. We consider a superconductor/ferromagnet/superconductor junction with a barrier of thickness $d$, as illustrated in Fig. 4(b). A finite phase difference $\phi$ exists between the two superconducting electrodes. At the left interface (position $-d/2$), the order parameter is given by $\Psi(-d/2) = u_1(x,y) \Psi_0$, and at the right interface ($d/2$), it is $\Psi(d/2) = e^{i\phi} u_2(x,y)\Psi_0$, where $\Psi_0$ denotes the amplitude of the order parameter deep within the superconducting bulk. The factors $u_1$ and $u_2$ encode the polarization or chirality of the superconducting states in the upper and lower layers, respectively. Within the ferromagnetic barrier, the order parameter can be approximated as a superposition of decaying and oscillatory components propagating from both interfaces. These represent the evanescent modes originating from the superconducting regions and describe how the coherence of the superconducting state penetrates and interferes within the junction.[40, 41]

$$\Psi(r) = Au_1(x,y)\exp(-\frac{z+d/2}{\xi_{F1}})\cos(\frac{z+d/2}{\xi_{F2}}) + Be^{i\phi}u_2(x,y)\exp(\frac{z-d/2}{\xi_{F1}})\cos(\frac{z-d/2}{\xi_{F2}}) \quad (5)$$

If $d < \xi_{F1,2}$, the tails of the superconducting wavefunctions from the left and right electrodes overlap significantly within the barrier region. This overlap gives rise to a finite Josephson coupling, resulting in a well-defined current–phase relation.

$$J(d/2) = 2img(\int dxdy \Psi^*(r)\frac{\partial}{\partial z}\Psi(r))$$

$$= \sin(\phi+\delta)|\int dxdy u_2^* u_1|\Psi_0^2[\frac{\frac{4}{\xi_{F1}}\exp(-\frac{d}{\xi_{F1}})\cos(\frac{d}{\xi_{F2}})}{\exp(-\frac{2d}{\xi_{F1}})\cos^2(\frac{d}{\xi_{F2}})-1} + \frac{\frac{2}{\xi_{F2}}\exp(-\frac{d}{\xi_{F1}})\sin(\frac{d}{\xi_{F2}})}{\exp(-\frac{2d}{\xi_{F1}})\cos^2(\frac{d}{\xi_{F2}})-1}] \quad (6)$$

Here, $\int dxdy u_2^* u_1 = e^{2i\gamma}|\int dxdy u_2^* u_1|$. Importantly, the geometric phase contributes to the interference in the same way as the normal phase.

$$J(d/2) \propto \sin(\phi + 2\gamma)|\int dxdy u_2^* u_1| \quad (7)$$

If the chirality of the left and right superconducting layers is opposite, the net Josephson current vanishes. This demonstrates that the geometric phase of the wavefunction can influence the tunneling current. As known, cuprate layers are known to exhibit superconductivity with predominantly $d_{xy}$-symmetry. Notably, $|\int dxdy u_2^* u_1| \propto \sin(2\theta)$, when both $u_1$ and $u_2$ are $d_{xy}$ symmetric. Experimentally, a dependence $\propto \sin(2\theta)$ of the tunneling current on the relative orientation (or polarization angle) between layers has been observed.[21] The comparison between experimental and theoretical results are shown in Fig. 4(c). A phase induced by the intrinsic magnetism is determined to be $9\pi/20$ by fitting experiments.

Focusing on the chiral degree of freedom, we find that superconducting states with opposite chirality acquire an additional, opposite geometric phase during tunneling through the vertical junction. The wavefunctions for the upper and lower layers are $\left| R_{\pm\theta/2} d_{xy} \right\rangle = \frac{1}{2i}(e^{-\pm i\frac{\theta}{2}} \left| l_z = 2 \right\rangle - e^{\pm i\frac{\theta}{2}} \left| l_z = -2 \right\rangle)$. The Josephson effect tell us $J \sim J_c \sin(l_z \gamma)$ for upper and lower layer are opposite for $l_z$=+2 and $l_z$=-2. The phase difference becomes opposite, resulting in cancellation of net charge current. However, even when the net tunneling current is zero in a pristine twisted junction, an orbital (chiral current) tunneling current can still persist, as illustrated in Fig. 4(b). If time-reversal symmetry is broken—either intrinsically by magnetic ordering or externally by a magnetic field—this balance is disturbed, and a net Josephson current arises, as shown in Fig. 4(d). The resulting current becomes chirally polarized. Thus, even if the superconducting ground state in twisted bilayer cuprates is not chiral, the system can act as a chiral filter, analogous to a spin filter in semiconductors.

In summary, we employed large-scale first-principles calculations to investigate the electronic structure of twisted bilayer $Bi_2Sr_2CaCu_2O_8$. Our results reveal that time-reversal symmetry can be spontaneously broken in the absence of topological superconductivity, and hidden flat bands emerge at the interface between atomic layers. We developed an effective model for twisted bilayer cuprates and, even without fully solving it, uncovered key physical insights—most notably, the emergence of a geometric phase analogous to the Pancharatnam-Berry phase in optics.

## ACKNOWLEDGMENTS

This work is supported by the